\documentclass{aa}

\usepackage[]{epsfig}
\def\nhf{{N$_{\rm H}^{f}$}}
\def\nhp{{N$_{\rm H}^{p}$}} 
\def\ltsima{$\; \buildrel < \over \sim \;$}
\def\simlt{\lower.5ex\hbox{\ltsima}}
\def\gtsima{$\; \buildrel > \over \sim \;$}
\def\simgt{\lower.5ex\hbox{\gtsima}}
\def\cgs{{erg cm$^{-2}$ s$^{-1}$}}
\def\ergs{{erg s$^{-1}$}}
\def\cm2{{cm$^{-2}$}}

\def\xddof{{$\chi^{2}/$dof}}

\def\xred{{$\chi^{2}_{\rm red}$}}
\def\xnn{{$\chi^{2}_{\rm \nu}$}}

\def\fhx{{$F_{\rm 2-10}$}}
\def\lum{{$L_{\rm 2-10}$}}

\def\p1{{Paper I}}

\def\xmm{{\em XMM--Newton}}
\def\chandra{{\em Chandra}}

\def\nhgal{{N$_{\rm H}^{\rm Gal}$}}
\def\nh{{N$_{\rm H}$}}

\def\chandra{{\em Chandra}}

\def\xmm{{\em XMM--Newton}}
\def\nhgal{{N$_{\rm H}^{\rm Gal}$}}
\def\nh{{N$_{\rm H}$}}

\def\pn{{\em PN}}
\def\mos{{\em MOS}}
\def\xr{{X--ray}}
\def\f14{{10$^{-14}$}}
\def\f13{{10$^{-13}$}}
\def\f12{{10$^{-12}$}}
\def\f11{{10$^{-11}$}}
\def\4u{{4U~1344$-$60}} 

\def\oiii{{[O\textsc{iii}] $\lambda$ 5007 \AA}}
\def\nii{{[N\textsc{ii}] $\lambda\lambda$ 6548,6583 \AA}}
\def\1nii{{[N\textsc{ii}] $\lambda$ 6548 \AA}}
\def\2nii{{[N\textsc{ii}] $\lambda$ 6583 \AA}}
\def\sii{{[S\textsc{ii}] $\lambda\lambda$ 6716,6731 \AA}}
\def\ssii{{[S\textsc{ii}]}}
\def\halpha{{H$_{\alpha}$}}
\def\nhalpha{{H$^{narrow}_{\alpha}$}}
\def\bhalpha{{H$^{broad}_{\alpha}$}}
\def\hbeta{{H$_{\beta}$}}

\begin{document}
 
\title{4U 1344$-$60: a bright intermediate Seyfert  galaxy \\
at $z$ =  0.012 with a relativistic Fe K$\alpha$ 
emission line\thanks{Based on observations obtained
with  XMM-Newton, an ESA Science Mission with instruments and
contributions  directly funded by ESA Member states and the USA
(NASA). Optical observations were carried out at the
European Southern Observatory, La Silla (Chile) under program No~
70.D-0227}}
 
\author{E.~Piconcelli\inst{1,2}, M.~S\'anchez-Portal\inst{2},
  M.~Guainazzi\inst{2},  A.~Martocchia\inst{3,4}, C.~Motch\inst{3}, 
A.~C.~Schr\"oder\inst{5}, S.~Bianchi\inst{2}, E. Jim\'enez-Bail\'on\inst{6},
G.~Matt\inst{6}}

\titlerunning{XMM-Newton observation of 4U 1344$-$60}
\authorrunning{E.~Piconcelli et al.}
 
\offprints{piconcelli@mporzio.astro.it} 
\institute{Osservatorio Astronomico di Roma (INAF), Via Frascati 33, I--00040
  Monteporzio Catone, Italy\and
European Space Astronomy Center (ESA), Apartado 50727, E--28080
Madrid, Spain\and Observatoire Astronomique de Strasbourg, rue de l'Universit\'e 11,
F--67000 Strasbourg, 
France\and Centre d'Etude Spatiale des Rayonnements, Avenue du Colonel
Roche, BP 4346, F--31028 Toulouse, France\and Department 
of Physics \& Astronomy (University of Leicester),
University Road, Leicester LE1 7RH, UK\and Dipartimento di Fisica
(Univerisit\`a degli Studi Roma Tre), Via della Vasca Navale 84,
I--00146 Roma, Italy}
 
\date{}
 
\abstract{We present analysis of the optical and X-ray spectra  of
  the low Galactic latitude bright (\fhx~= 3.6  $\times$
10$^{-11}$ \cgs) source \4u. 
On the basis of the optical data
we propose to classify \4u~as an  intermediate type Seyfert galaxy and 
we measure a value of $z$ = 0.012$\pm$0.001 for its redshift.
From the \xmm~observation we find that the overall X--ray
spectral shape of \4u~is complex and can be described 
by a power-law continuum ($\Gamma$ $\approx$ 1.55) obscured by two
  neutral absorption components (\nhf~$\sim$ 10$^{22}$ \cm2 and
  \nhp~$\sim$ 4 $\times$ 10$^{22}$
\cm2), the latter covering only  the $\sim$ 50\% of the primary X-ray source.
The X-ray data therefore lend support to our classification of \4u. It
  exhibits a broad and skewed Fe K$\alpha$ line at $\sim$ 6.4 keV,
 which suggests the existence of an
accretion disk that is able to reprocess the primary continuum  down to
a few gravitational radii. Such a line represents one of the clearest
  examples of a relativistic line observed by \xmm~so far.
Our analysis has also revealed the  marginal presence of two narrow line-like emission
  features at $\sim$ 4.9 and $\sim$ 5.2 keV.

 }
 
\maketitle
 
\section{Introduction}

\4u~was discovered by the {\it Uhuru} (SAS A) \xr~Observatory when scanning
of the Galactic plane (Forman et al. 1978). Wood et al. (1984) reported a flux of
$\sim$ 2 $\times$ \f11~\cgs~in the 2--10 keV band for this \xr~source 
in the Large Area Sky Survey performed with {\it HEAO-1}.
The first position of \4u~with good astrometric accuracy has been provided by 
Warwick et al. (1988)
using {\it EXOSAT} data (R.A.(B1950) =13$^{h}$ 43$^{m}$ 59$^{s}$.5, 
Dec.(B1950)=$-$60$^{\circ}$ 23$^{\prime}$ 39$^{\prime\prime}$).
Because \4u~lies deeply in the
Galactic plane ($l = 309\fdg77, b = 1\fdg51$),
it is not easily accessible at all wavelengths
(in particular the optical to soft X--ray bands). For this reason,
\4u~has been optically identified only very
recently by Masetti et al. (2006;  see also Michel et al. 2004) 
 as a Type 1 Seyfert galaxy at $z$ = 0.013$\pm$0.001.
Despite its X-ray brightness, no \xr~satellite directly
pointed \4u~ with an adequate exposure time, so its X--ray spectral
properties were almost unknown up to \xmm.

This source is included in the first {\it INTEGRAL} AGN
Catalog (flux of $\sim$ \f11~\cgs~in the 20--100 keV band), 
which has been recently published by Beckmann et al. (2005).
They performed a first-order spectral analysis fitting  \xmm, {\it INTEGRAL} ISGRI,
and SPI data, which turned out a best--fit model consisting of an absorbed (\nh~=
 2.64$\pm$0.07 $\times$ 10$^{22}$ \cm2) power law with a photon index $\Gamma$ =
 1.65$^{+0.02}_{-0.03}$ plus a Gaussian line at 6.13$^{+0.08}_{-0.09}$ keV.

In this paper we report the identification and the spectrum of the optical counterpart
of this bright high--energy source (Sect.~2) and
present the results of a detailed analysis of its \xmm~spectrum
(Sects.~3 and 4). 
In Sect.~5 we discuss our findings providing clues to the nature and the origin of the
different features observed in the X--ray spectrum and their connection with the
optical classification of \4u.
Finally, in Sect.~6, we summarize the results of the X--ray and optical analysis
and give our conclusions.

\section{The optical properties of \4u}
\label{sec:optical}
The optical imaging and spectroscopic observations of
\4u~(P.I. C. Motch)
were carried
 out on \mbox{01 Mar 2003} with the EFOSC 2 imaging spectrograph at
 the Cassegrain focus of the ESO 3.6m telescope at La Silla (Chile),
% as part of the Program ID 70.D-0227
 in the framework of the Galactic plane part of the identification
program of the \xmm~Survey Science Center (Motch et al. 2003; Watson
et al. 2001). 
A thinned  Loral/Lesser CCD detector, providing a  2060 $\times$ 2060~pixel
 image was used. Pixel size is 15 $\mu$m, resulting into a plate scale
 of 0$^{\prime\prime}$.158/pixel.   A \mbox{ 2 $\times$ 2 } binning
 was applied in both imaging and spectroscopical observations.
These data are the same as those analyzed by Masetti et al. (2006).

\begin{figure}
\begin{center}
\centerline{
\psfig{figure=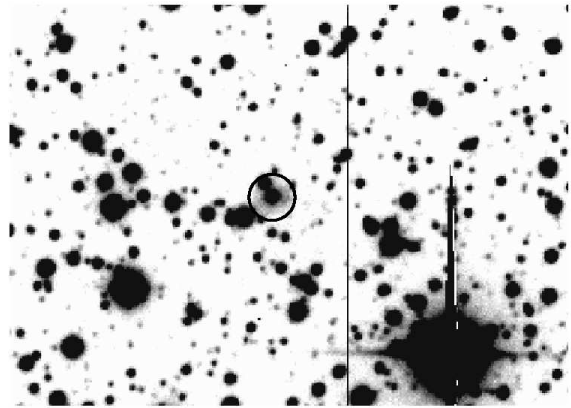,width=7.5cm}}
\end{center}
\caption{Gunn {\em i} image of the field of \4u.~North is up and East to the left.
A circle of r = 4$^{\prime\prime}$ encloses the two candidates to be the optical counterpart
of the \xmm~source. Based on the analysis of our optical spectra, 
we determined
that the counterpart of \4u~is the source located to the SW. The signature of the host galaxy is 
clearly seen as a faint nebulosity in the figure.}
\label{fig:4U_I}
\end{figure}

A 300~s image (Fig. \ref{fig:4U_I}) was obtained through the ESO
Gunn {\em i} \#705 filter.  
 Images were corrected for bias and flat-field using sky exposures
obtained at dawn and with standard MIDAS procedures.
 The seeing value derived from this image is \mbox{FWHM $\simeq$
1$^{\prime\prime}$}. 
Absolute astrometry was performed using the
USNO B1 catalogue. According to  our field astrometry, the
  coordinates of the  object  proposed by Masetti et al. (2006) as
  the optical counterpart of the X--ray source
 are
R.A.(J2000) = 13$^{h}$ 47$^{m}$ 36$^{s}$.01 and
Dec.(J2000) = $-$60$^{\circ}$ 37$^{\prime}$ 03$^{\prime\prime}$.75 
(see Fig.~\ref{fig:4U_I} and comments therein), in
excellent agreement with our estimate of the position of the
\xmm~source (R.A.(J2000) = 13$^{h}$ 47$^{m}$ 36$^{s}$.1 and
Dec.(J2000) = $-$60$^{\circ}$ 37$^{\prime}$ 03$^{\prime\prime}$.3).
These coordinates also match (to a few tenths of arcsec) those of
the source \#45 (IRSF J134736.0-604704) 
in the deep NIR survey 
carried out by  Nagayama et al. (2004). The object was identified as a spiral galaxy.
Furthermore, Schr\"oder et al. (2006; see also Schr\"oder et al. 2005)
cross-identified their galaxy DZOA4653-11 with 4U 1344-60 and found it to
be a member of a group or small cluster around the giant radio galaxy
Centaurus B. They measured a $K_s$-band magnitude of $10\fm56$, which is
$9\fm48$ after correction for Galactic extinction according to the maps by
Schlegel et al. (1998).

%%%%%%%%%%%%%%%%%%%%%%%%%%%%%%%%%%%%%%%%%%%%%%%%%%%%%%%%%%%%%%%%%%%%%%%%%%
%%%%%%%%%%%%%%%%%%%%%%%%%%%%%%%%%%%%%%%%%%%%%%%%%%%%%%%%%%%%%%%%%%%%%%%%%%
A 1000~s spectrum was obtained using the ESO grism \#06, providing
a  resolution of \mbox{12.9 \AA~(FWHM)}  and a dispersion of
\mbox{4.12 \AA~per binned pixel}. A 1$^{\prime\prime}$.5-wide slit was
used. 
All CCD spectroscopic frames were corrected for bias and flat-fielded
using standard MIDAS procedures. The wavelength calibration was derived
from the observation of He $+$ Argon arc lamps. One-dimensional spectra
were extracted with a procedure optimizing the accumulation region in
order to reach the best signal-to-noise and using sky background cleaned
from cosmic-ray impacts. All spectra were corrected for atmospheric
extinction and calibrated in flux using spectrophotometric standards.

\begin{figure*}
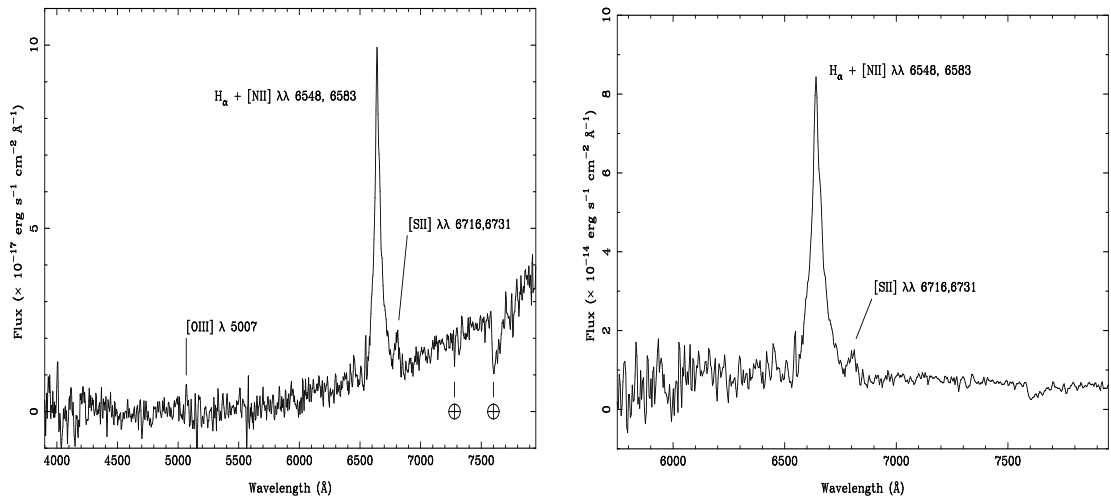

\begin{center}
\centerline{
\psfig{figure=ep4u_f2.ps,height=7.cm,width=6.5cm,angle=-90}\hspace{0.5cm}
\psfig{figure=ep4u_f3.ps,height=7.cm,width=6.5cm,angle=-90}}
\end{center}
\caption{Optical spectrum of \4u~before ({\it left}) and after ({\it
    right}) applying the reddening correction. For the latter  
the wavelength range has been restricted to $\lambda >$  5750
\AA~since the bluer part of the spectrum is affected by large noise.
The features marked with the $\oplus$ symbol are identified as atmospheric
telluric bands. Prominent emission lines are marked for $z$ = 0.012.}
\label{fig:4U_opt_spec}
\end{figure*}

The one-dimensional, wavelength,
and flux-calibrated spectrum  of the optical counterpart of the X-ray
source is shown in Fig.~\ref{fig:4U_opt_spec}. The \halpha~line
complex is very intense.   The spectrum is affected by a large Galactic
extinction and the continuum disappears short-wards $\sim$ 5750
\AA. Nevertheless, a trace of the strong  \oiii~feature was still
detected.   The spectrum was dereddened using the extinction  curve
from Cardelli et al. (1989), setting $E(B-V) = 2.93$ mag according to
Schlegel et al. (1998). While the \oiii~line became more evident, the
resulting  spectrum was extremely noisy below $\sim$ 5750
\AA. Therefore we concentrated our analysis on the red portion of the
spectrum.

%%%%%%%%%%%%%%%%%%%%%%%%%%%%%%%%%%%%%%%%%%%%%%%%%%%%%%%%%%%%%%%%%%%%%%%
%%%%%%%%%%%%%%%%%%%%%%%%%%%%%%%%%%%%%%%%%%%%%%%%
The main spectral feature observed is the \halpha+\nii~line
complex. It is roughly symmetric, peaking at the
\nhalpha~component.  To compute the galaxy  redshift, the line complex
core was fitted by means of a single Gaussian profile. From the
computed central wavelength, a value of  \mbox{z = 0.012 $\pm$ 0.001}
was derived.  An attempt to deblend the four components of the
complex (\bhalpha,~\nhalpha,~\1nii, \& \2nii) was performed. To
this end,  the observer's frame expected central wavelengths of the
different lines have been computed based on our redshift estimate.
Then, a simultaneous fit of four Gaussian profiles was performed. Even
though it was not possible to perform a completely  stable fit of
the four components, the procedure allowed us  to roughly 
estimate the flux of the two main components:  I(\nhalpha) =
9.3 $\pm$ 1.5 $\times$ 10$^{-13}$ \cgs~and I(\bhalpha)= 3.3 $\pm$
0.8$\times$ 10$^{-12}$ \cgs.~The flux in  the \sii~doublet was  also
estimated as $\simeq$ 3.0 $\times$ 10$^{-13}$ \cgs. Therefore we
obtained a ratio \nhalpha/\ssii~$\simeq$~3,  consistent with an AGN
spectrum (Veilleux \& Osterbrock 1987). 
The width of the broad component of \halpha~was $\sim$ 4400
km s$^{-1}$ FWHM, a typical value for  a Seyfert 1.x galaxy.

Moreover, even though the \hbeta~line was not detected in our spectrum,
we could derive an estimate of the Seyfert type that was more accurate
 than the one provided by Masetti et al. (2006)
using  the  approximate relation from
Netzer (1990): {\it Seyfert Type} 
$\simeq 1 +\left[I(\mbox{\nhalpha})/I(\mbox{\bhalpha})\right]^{0.4}$. 
Using this
relation and the flux values given above, we propose to classify
\4u~as a Seyfert 1.5 galaxy. However, it should be taken into account
that the large flux uncertainties could shift this tentative
classification close to the Seyfert 1.8 class.

\section{XMM-Newton observation and data reduction}

On August 25, 2001 \4u~was serendipitously observed by \xmm~(Jansen et
al. 2001 and references therein) for 37 ks (Obs. Id. 0092140101). 
This source lies at $\sim$14 arcmins
from the radio-galaxy Centaurus B, which was the target of the
observation.

\xmm~data were processed with SAS v6.5.  The  EPCHAIN and  EMCHAIN 
tasks were used for processing the raw \pn~and \mos~data files,
respectively, and for generating the relative linearized event files.
X--ray events corresponding to patterns 0--12(0--4)  for the
\mos(\pn)~cameras were selected. We employed the most updated
calibration files at the time the reduction was performed (September
2005). \4u~was found to be well within the \pn~field-of-view (FOV)
while it was outside the {\it MOS2} FOV and just on the edge of the
{\it MOS1} FOV. Given  the current calibration uncertainties at 
such an extreme off-axis location,  {\it MOS1} data were ignored in our
analysis.  The \pn~event list was furthermore filtered to ignore
periods of high  background flaring  according to the method presented
in Piconcelli et al. (2004) based on the cumulative distribution
function of background lightcurve count-rates.  After this data
cleaning, we obtained a net exposure time of 25.5 ks.  The spectrum
was binned to a minimum of 35 counts per bin to apply  the $\chi^{2}$
minimization technique in the spectral fitting, which was performed in
the 0.3--10 keV band using the XSPEC package (v.11.3).

The quoted errors on the model parameters correspond to a 90\%
confidence level for one interesting parameter ($\Delta\chi^2$ = 2.71;
Avni 1976). All luminosities are calculated assuming a $\Lambda$CDM
cosmology with ($\Omega_{\rm M}$,$\Omega_{\rm \Lambda}$) = (0.3,0.7)
and a Hubble constant of 70 km s$^{-1}$ Mpc$^{-1}$ (Bennett et
al. 2003).
%%%%%%%%%%%%%%%%%%%%%%%%%%%%%%%%%%%%%%%%%%%%%%%%%%%%%%%%%%%%%%%%%%%%%%%
\begin{figure*}
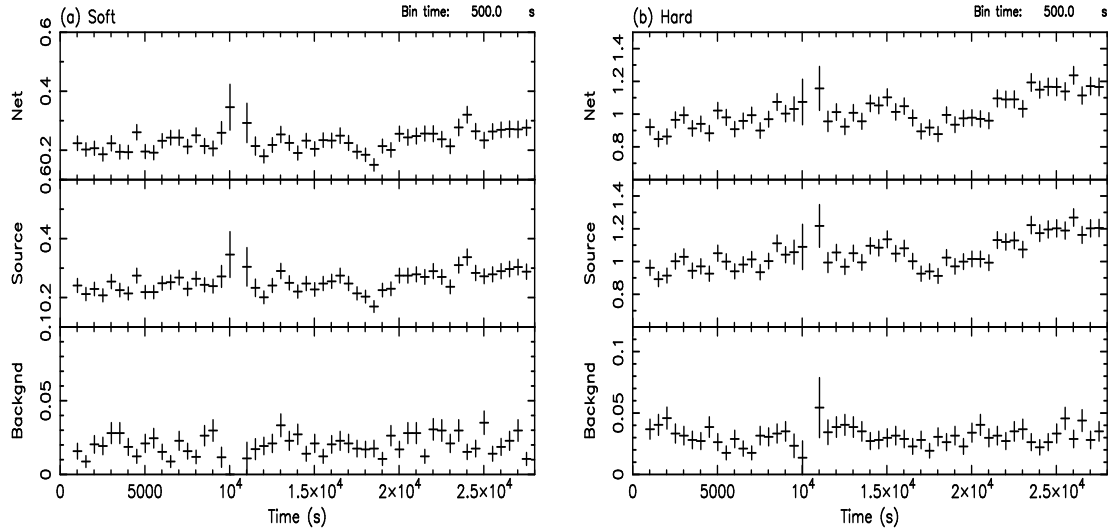

\begin{center}
\centerline{
\psfig{figure=ep4u_f4.ps,height=7.cm,width=7.0cm,angle=-90}\hspace{0.5cm}
\psfig{figure=ep4u_f5.ps,height=7.cm,width=7.0cm,angle=-90}}
\end{center}
\caption{Lightcurves for the \xmm~observation of \4u~in the soft 0.2--2 keV ({\it left})
 and  hard 2--15 keV ({\it right}) band ({\it Upper}: background
 subtracted lightcurve; {\it Middle}: source$+$background lightcurve ; {\it Lower}:
 lightcurve for the background, with count rates rescaled to the area
 of the source extraction region).}
\label{fig:lc}
\end{figure*}
%%%%%%%%%%%%%%%%%%%%%%%%%%%%%%%%%%%%%%%%%%%%%%%%
%%%%%%%%%%%%%%%%%%%%%%%%%%%%%%%%%%%%%%%%%%%%%%%%%%%%%%%%%%%%%%%%%%%%%%%%%%%%
%%%%%%%%%%%%%%%%%%%%%%%%%%%%%%%%%%%%%%%%
%%%%%%%%%%%%%%%%%%%%%%%%%%%%%%%%%%%%%%%%%%%%%%%%%%%%%%%%%%%%%%%%%%%%%%%%%%
%%%%%%%%%%%%%%%%%%%%%%%%%%%%%%%%%%%%%%%%%%%%%%%%%%%%%%%%%%%%%%%%%%%%%%%%%%%
%%%%%%%%%%%%%%%%%%%%%%%%%%%%%%%%%%%%%%%%%%%%%%%%%%%%%%%%%%%%%%%%%%%%%%%%%
%%%%%%%%%%%%%%%%%%%%%%%%%%%%%%%%%%%%%%%%%%%%%%%%%%%%%%%%%%%%%%%%%%%%%%%%%%%
%%%%%%%%%%%%%%%%%%%%%%%%%%%%%%%%%%%%%%%%%%%%%%%%%%%%%%%%%%%%%%%%%%%%%%%%%
\section{\xmm~results}
\subsection{Temporal analysis}

X--ray light curves were extracted from the \pn~dataset in the
soft (0.2--2 keV) and hard (2--15 keV)  bands. The time bin size was
set to be 500 s for all light curves.  Background light curves were
extracted from source-free regions in the same energy bands and with
the same bin sizes.  In Fig.~\ref{fig:lc} the
background--subtracted, source$+$background and  background light
curves in the soft and hard band are plotted. Note that the datapoints
belonging to the final interval of the observation (i.e. data taken
at times $t_{\rm obs} >$ 28 ks after the start of the observation)
were excluded as they are completely dominated by a large flaring
particle background. 

 A positive 30--35 per
cent variation in count rate (CR) in both the soft and the hard bands 
occurred at $t_{\rm obs}$ \simgt~20 ks.
The hardness ratio  (CR$_{\rm Soft}/$CR$_{\rm Hard}$) thereby
remained steady, within the errors, during the whole portion of
 the observation selected for the spectral analysis.

\subsection{Spectral analysis}
\label{sec:analysis}
\begin{figure*}
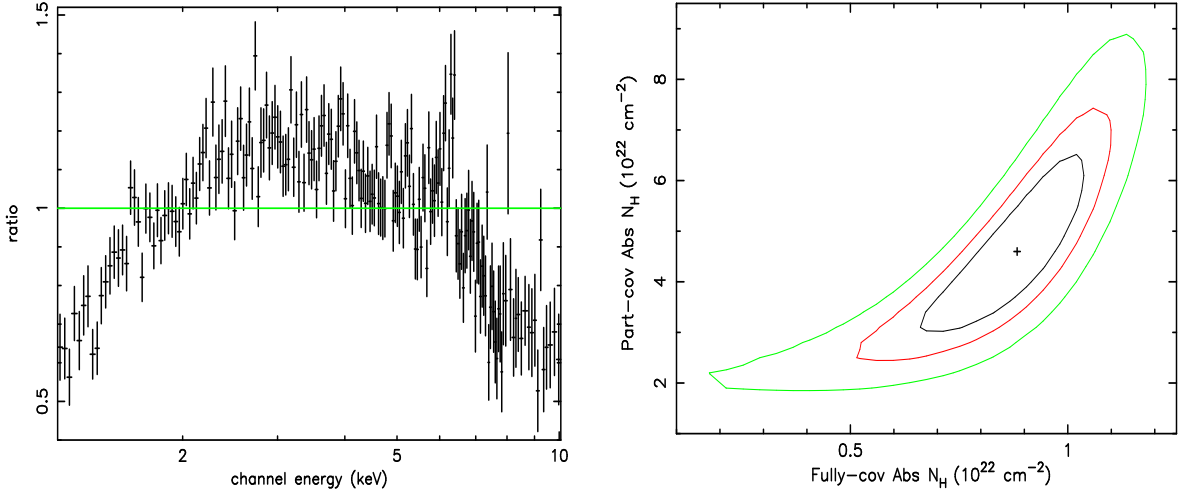

\begin{center}
\centerline{
\psfig{figure=ep4u_f6.ps,height=7.5cm,width=6.5cm,angle=-90}\hspace{0.5cm}
\psfig{figure=ep4u_f7.ps,height=7.5cm,width=6.5cm,angle=-90}}
\end{center}
\caption{
{\it Left:} (a) Data to model ratios for a model comprising
  Galactic absorption plus power law. {\it Right:} (b) 
Confidence contour plot showing the column density  of the 
fully-covering absorber against the column density of 
the partially-covering absorber obtained by applying model A. 
The contours are at 68\%, 90\%, and 99\%  confidence levels for two interesting parameters, respectively.}
\label{fig:cont}
\end{figure*}

All the models presented in this section include a large
absorption component due to the line--of--sight Galactic column of
\nh~= 1.08 $\times$ 10$^{22}$
\cm2~(Dickey \& Lockman 1990). If not specified, values of the
physical parameters are
reported in the source frame.
As shown in Fig.~\ref{fig:cont}a, the X-ray spectrum of \4u~is
complex, and a simple fit with a power law  ($\Gamma
\approx$ 0.7)
 yielded a very
poor fit with \xnn~= 4.7. The most prominent features in the
residuals are the low energy cut off and the positive excess at
$\sim$6 keV. The latter is readily explained in terms of
fluorescent emission from the iron K-shell as commonly observed in
most Seyfert galaxies, while the   negative residuals below 2 keV
and  the extremely flat slope are indicative of heavy absorption along
our line of sight. Due to the  spectral complexity of
\4u, we initially excluded the data from 5 to 7 keV to better
determine the underlying continuum.  

An absorbed power law  yielded a marginally acceptable fit to the 
data with an associated \xddof~=177$/$147
(i.e. \xnn~$\approx$1.2) and a still unusually flat $\Gamma \sim$  1.3, which suggests the
presence of more complex absorption.  We
then applied a partial-covering model to the data, which
usually provides a good fit in the case of absorbed 
  intermediate-type Seyfert 
galaxies (e.g. 
Immler et al. 2003; Schurch \& Warwick 2003; Pounds et al. 2004).
This parameterization, consisting of two power
laws with only the second one passing through the absorber
and with the two photon indices assumed to be identical,
produced a decrease of
$\Delta\chi^2$ = 29 compared with the  absorbed power-law model, i.e. a statistical
improvement at $>$99.9\% confidence level according to an $F$--test.
This fit resulted in a photon index of $\Gamma$ = 1.46$\pm$0.06 and a column density
of the partial covering absorber of \nhp~= 2.0$\pm$0.3 $\times$ 10$^{22}$ \cm2.
 
 %%%%%%%%%%%%%%%%%%%%%%%%%%%%%%%%%%%%%%%%%%%%%%%%%%%%%%%%%%%%%%%%%%%%%%%%%%
%%%%%%%%%%%%%%%%%%%%%%%%%%%%%%%%%%%%%%%%%%%%%%%%%%%%%%%%%%%%%%%%%%%%%%%%%%
%%%%%%%%%%%%%%%%%%%%%%%%%%%%%%%%%%%%%%%%%%%%%%%%%%%%%%%%%%%%%%%%%%%%%%%%%%

Furthermore,  we found that the addition of a fully-covering intrinsic
absorption component gave a further, highly significant (at $>$99.9\%
c. l.) improvement to the fit statistics, with \xddof~= 138/145.
%%%%%%%%%%%%%%%%%%%%%%%%%%%%%%%%%%%%%%5
\begin{table*}
\caption{Results of the spectral analysis.}
\label{tab:results}
\begin{center}
\begin{tabular}{ccccccccc}
\hline\hline
 \multicolumn{1}{c} {Model}& \multicolumn{1}{c} {$\Gamma$}&
\multicolumn{1}{c} {N$_{\rm H}^{f}$ (Fully-cov.)}& \multicolumn{1}{c}
{\nhp~(Part.-cov.)}& \multicolumn{1}{c} {$E_{\rm Fe}$} &
\multicolumn{1}{c} {R$_{\rm in}$/$\sigma_{\rm Fe}$} &
\multicolumn{1}{c} {EW} & \multicolumn{1}{c} {$i$} &
\multicolumn{1}{c} {\xddof}\\ 
 &          &  10$^{22}$ \cm2
& 10$^{22}$ \cm2& keV        &$R_g$/keV         &eV & deg&\\
\hline\hline\\
A$^a$&1.59$^{+0.11}_{-0.10}$&0.88$^{+0.17}_{-0.25}$&4.61$^{+2.11}_{-1.75}$
&$-$&$-$&$-$&$-$&138/145\\
B&1.57$^{+0.03}_{-0.11}$&0.69$^{+0.10}_{-0.51}$&3.94$^{+1.79}_{-0.76}$&
6.18$^{+0.07}_{-0.07}$ &0.27$^{+0.09}_{-0.09}$&393$^{+122}_{-107}$&$-$&175/179\\
C&1.54$^{+0.09}_{-0.10}$&0.69$^{+0.26}_{-0.39}$&3.90$^{+2.15}_{-0.92}$&
6.47$^{+0.04}_{-0.03}$&14$^{+4}_{-4}$&378$^{+87}_{-81}$&$\leq$15&167/178 \\ 
D&1.54$^{+0.09}_{-0.08}$ &0.68$^{+0.26}_{-0.51}$ &3.83$^{+1.97}_{-0.63}$
&6.28$^{+0.11}_{-0.09}$&$\geq$13&383$^{+52}_{-40}$ &26$^{+23}_{-12}$&170/178\\ 
E& 1.69$^{+0.02}_{-0.03}$& 1.02$^{+0.04}_{-0.04}$&7.59$^{+0.60}_{-1.34}$
&6.17$^{+0.04}_{-0.04}$&0.0f.$^b$&96$^{+30}_{-36}$&$-$&180/178\\
 &         &                   &
&6.42$^{+0.27}_{-0.04}$&0.0f.$^b$&155$^{+40}_{-40}$&$-$&\\ \hline
\end{tabular}
\end{center} 
Model A: absorbed partial-covering; Models B, C, and D include the same
underlying continuum as Model A plus: a broad Gaussian line (B); a 
Schwarzschild {\tt diskline} (C);  a Kerr {\tt laor} line (D) and two
narrow Gaussian lines to account for the Fe K$\alpha$ emission.
See text for  further details of
  each model. 
$^a$ Data from 5 to 7 keV were excluded. $^b$ Value of the line width
  $\sigma_{\rm Fe}$  fixed to 0 keV.
\end{table*}
%%%%%%%%%%%%%%%%%%%%%%%%%%%%%%%%%%%%%%%%%% $^{+}_{-}$
This model (e.g. model A in Table 1) yielded 
a photon index of $\Gamma$ = 1.59$^{+0.11}_{-0.10}$ and a
column density of \nhf~=  0.88$^{+0.17}_{-0.25}$ $\times$
10$^{22}$ \cm2~and \nhp~= 4.61$^{+2.11}_{-1.75}$ $\times$ 10$^{22}$
\cm2~for the absorber fully--covering and partially--covering
(covering--fraction C$_f$ = 43$\pm$10\%) the source, respectively. 
The two--dimensional contour plot in the parameter space \nhf--\nhp~is shown in Fig.~\ref{fig:cont}b. 
The null hypothesis
probability associated to this fit, $P$ = 0.66,  and the visual
inspection of the data-to-model ratios suggested that it
provides an appropriate description of the broad-band X-ray spectral
shape of \4u. 

 Moreover, we also fitted the X--ray continuum of \4u~using an alternative
  model consisting of an absorbed power law including a reflection component 
({\tt pexrav} model in XSPEC), which is expected due to the presence
  of a strong Fe K$\alpha$ line. However, we yielded a significantly ($>$99.9\%) worse fit than model A
  with a final \xred~= 1.21 for 146 d.o.f. The photon index was flat
  ($\Gamma$ $\sim$ 1.3) and we measured only an upper limit for the
  strength of the reflection ($R$ \simlt~0.5).

%%%%%%%%%%%%%%%%%%%%%%%%%%%%%%%%%%%%%%%%%%%%%%%%%%%%%%%%%%%%%%%%%%%%%%%%%%
\begin{figure*}
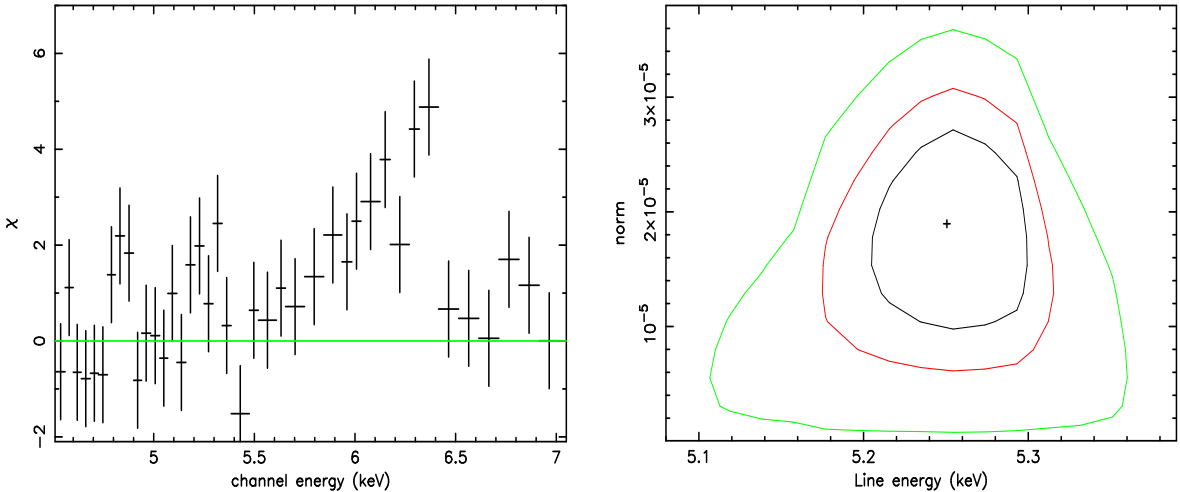

\begin{center}
\centerline{
\psfig{figure=ep4u_f8.ps,height=7.5cm,width=6.5cm,angle=-90}\hspace{0.5cm}
\psfig{figure=ep4u_f9.ps,height=7.5cm,width=6.5cm,angle=-90}}
\end{center}
\caption{
{\it Left:}~(a)~Close-up of the deviations, in units of sigma, in the 4.5--7
  keV band of the
  observed EPIC \pn~data from model A, i.e.  an absorbed
  partial
covering model fitted to the (0.3--5 keV)$+$(7--10 keV) range (see
  Table~1) and extrapolated to the whole \pn~energy interval.
{\it Right:}~(b)~Confidence contour plot showing the energy against
the photon flux (in units of ph \cgs)
of the line detected at $\sim$ 5.2 keV.   
The contours are at 68\%, 90\%, and 99\%
 confidence levels for two interesting parameters.}
\label{fig:popup}
\end{figure*}
%%%%%%%%%%%%%%%%%%%%%%%%%%%%%%%%%%%%%%%%%%%%%%%%%%%%%%%%%%%%%%%%%%%%%%%%%%
Figure~\ref{fig:popup} shows the residuals obtained by
extrapolating the model A fit above the energy band between 5 and 7
keV, which was previously ignored. A large excess is evident from
$\sim$ 5.5 to $\sim$ 6.5 keV, indicating that the line profile
is broad and asymmetrical with the peak of the emission located around
6.4 keV. Furthermore,  this figure suggests the 
presence of two
emission line-like features at $\sim$ 4.8 and $\sim$ 5.2 keV.

We started to fit the  complex Fe K$\alpha$ emission
with a  Gaussian line. 
This fit (labelled as model B in Table~1) had an associated
\xddof~=175(179)
and yielded an improvement at $>$ 99.99\%
confidence level according to an F-test once compared with a fit
similar to model A but with the whole 0.3--12 keV band taken into
account.
The line was at 6.18$\pm$0.07 keV and turned out to be both strong (equivalent width EW =
393$^{+122}_{-107}$ eV) and
spectrally resolved with a $\sigma_{\rm Fe}$ = 0.27$\pm$0.09 keV.
If due to Doppler shift, the corresponding velocity width would be
$\sim$ 30000$\pm$10000 km~s$^{-1}$, a value that is typical of the innermost
regions of the accretion disk around a supermassive black hole (BH) and
only three times lower than the extreme ($\sim$ 10$^5$ km~s$^{-1}$) 
relativistic Fe K$\alpha$
line observed in MGC-6--30-15 (Tanaka et al. 1995).
Interestingly, a similar value of the velocity FWHM, i.e. $\sim$ 30000
km~s$^{-1}$, has also been found for most of the broad iron lines
detected by \xmm~so far (e.g. Porquet \& Reeves 2003; Balestra et
al. 2004; Porquet 2006).

We therefore fitted the skewed Fe K$\alpha$ emission line in \4u~with  
 the broad profile expected
from a relativistic accretion disk around: ($i$) a non-rotating
(Schwarzschild)  BH (model {\tt diskline} in XSPEC; Fabian et
al. 1989), ($ii$)  a spinning (Kerr) BH
(model {\tt laor} in XSPEC; Laor 1991). An emissivity law  with $q$ fixed to
the customary value of 2.5  was assumed (Nandra et al. 1997).
These spectral fits are indicated in Table~1 as models
C and D. We also applied an alternative model consisting of two 
narrow Gaussian lines (model E). 
 
Both fits with a relativistic line profile provided an excellent
description of the Fe line observed in \4u~ with an associated
\xred~= 0.94 (model C) and
\xred~= 0.95 (model D) for 178 degrees of freedom. 
The best-fit values
of the spectral parameters of the line are reported in Table~1.
In both cases we found a disk inclination \simlt~45 degrees and
a disk inner radius larger than the marginally
stable orbit in a Schwarzschild/Kerr metric.

It is worth noting that assuming a different underlying continuum model
(i.e. {\tt pexrav} as in Sect.~4) did not modify the diskline parameters.
In fact, we found similar values ($E$ = 6.38$\pm$0.06 keV, $R_{\rm in}$
14$^{+5}_{-4}$ R$_g$) to those found with model C, with the
obvious exception of the EW, which decreased to $\sim$ 150 eV.

Finally, we did not find any statistical requirement for a narrow Fe
K$\alpha$ line at 6.4 keV when a relativistic line model was applied.
A narrow Fe line is expected to emerge from the absorbing
material. When \nh~is a few $\times$ 10$^{22}$ \cm2, as
observed in \4u, the predicted EW of this emission line is about 40
eV (Makishima 1986), which is  consistent with the upper limit of 
EW $\leq$ 39 eV we inferred from the spectrum.

The application of model E (e.g. Table~1 for details) to the data also
produced a statistically
good fit. The energies(EWs) of the two narrow lines were 
6.17$^{+0.04}_{-0.04}$(EW = 96$^{+30}_{-26}$) and
6.42$^{+0.27}_{-0.04}$(EW = 155$\pm$40 eV) keV.
 However, the final $\Delta\chi^2$ = 13 
with respect to model C and the presence of an excess in the
data--to--model residuals redward of the 6.1 keV line led us to
reject this model on statistical grounds.

 Models from B to E yielded a value of the covering factor
of the partially--covering absorber C$_f$ = 50$\pm$15\%.
Once Model C was assumed, we measured a flux \fhx~= 3.6  $\times$
10$^{-11}$ \cgs~in the hard (2--10 keV) energy band that corresponds,
after the correction for both Galactic and intrinsic absorption
components, to a luminosity  \lum~= 1.48 $\times$
10$^{43}$ \ergs.\\

 Some positive excesses were present
in the data--to--model ratio residuals: apart the two features around
$\sim$ 5 keV shown in  Fig.~\ref{fig:popup}a, a third one was found at $\sim$ 1.6 keV.
All previous models therefore also included three Gaussian lines to reproduce 
these spectral features. Each line represents an improvement in the resulting fit statistics
significant at $\geq$95.5\% confidence level. Table~2 lists the line's
best-fit parameters, while Fig.~\ref{fig:popup}b shows the two-dimensional
  iso-$\chi^2$ contour plot in the parameter space $E$--Photon flux for the
  line detected (98.5\% c.l.) at 5.31$\pm$0.05 keV.

 The line at 1.6 keV is puzzling since no obvious strong emission
  line is expected to emerge at these energies. 
We therefore checked the possibility that such a feature could be an 
artifact owing to calibration uncertainties in the \pn~instrumental
response function at  the extreme off-axis
position of \4u. The energy range between 1 and 2 keV is rich in
emission lines and edges due to the detector materials.
In particular, the effective area of the \xmm~mirror shows a pronounced
edge near 2 keV (the Au M edge), and  the \pn~quantum efficiency shows a prominent
feature at 1.8 keV caused by the Si K edge (Str\"uder et al. 2001).
In the background spectrum,  two
strong emission lines are present at $\sim$ 1.5 keV (Al K$\alpha$ line
in the detector background) and  $\sim$ 1.75 keV (Si K$\alpha$).
It is thus likely that the feature observed at 1.6 keV may be due to 
uncertainties in the response matrix and background subtraction at
the  source's off-axis location 
and, therefore,
will be ignored in the following discussion.
On the other hand, the lines detected around 5 keV are not affected 
by this problem since in this energy band the response of the detector
is flat, and no strong spectral features are reported in the background spectrum.

%%%%%%%%%%%%%%%%%%%%%%%%%%%%%%%%%%%%%%%%%%%%%%%%%%%%%%%%%%%%%%%%%%%%%%%%%%%%
\begin{table}
\caption{Best-fit value of the parameters of the three narrow emission lines detected in the spectrum.}
\label{tab:lines}
\begin{center}
\begin{tabular}{ccc}
\hline\hline
 \multicolumn{1}{c} {Energy}& \multicolumn{1}{c} {EW}&
\multicolumn{1}{c} {$P$(F--test)}\\ 
keV&eV          & \% \\
\hline\hline\\
1.63$\pm$0.02&19$^{+10}_{-12}$ &98.5\\
4.90$\pm$0.04&45$^{+25}_{-22}$ &95.5\\
5.31$\pm$0.05&57$^{+27}_{-30}$ &98.5\\
\hline
\end{tabular}
\end{center}
\end{table}
%%%%%%%%%%%%%%%%%%%%%%%%%%%%%%%%%%%%%%%%%%%%%%%%%%%%%%%%%%%%%%%%%%%%%%%%%%%%
%%%%%%%%%%%%%%%%%%%%%%%%%%%%%%%%%%%%%%%%
%%%%%%%%%%%%%%%%%%%%%%%%%%%%%%%%%%%%%%%%%%%%%%%%%%%%%%%%%%%%%%%%%%%%%%%%%%
%%%%%%%%%%%%%%%%%%%%%%%%%%%%%%%%%%%%%%%%%%%%%%%%%%%%%%%%%%%%%%%%%%%%%%%%%%%
%%%%%%%%%%%%%%%%%%%%%%%%%%%%%%%%%%%%%%%%%%%%%%%%%%%%%%%%%%%%%%%%%%%%%%%%%
%%%%%%%%%%%%%%%%%%%%%%%%%%%%%%%%%%%%%%%%%%%%%%%%%%%%%%%%%%%%%%%%%%%%%%%%%%%
%%%%%%%%%%%%%%%%%%%%%%%%%%%%%%%%%%%%%%%%%%%%%%%%%%%%%%%%%%%%%%%%%%%%%%%%%
\section{Discussion}

\subsection{Optical classification and X-ray absorption}

According to our analysis in  Sect.~\ref{sec:optical}, \4u~is a
Seyfert 1.5 at $z$ = 0.012$\pm$0.001. However,
given the large uncertainties in the
estimates of the \bhalpha~and \nhalpha~fluxes, a Seyfert 1.8
classification could also be possible.
We therefore only partially confirmed the results of Masetti et al. (2006), as they
report a Seyfert 1  optical classification.
From the \xmm~observation we found that the overall X--ray
spectral shape of \4u~is complex and can be described 
by a power-law continuum modified by two neutral absorption components, one of
which covers only  50\% of the primary X-ray source.
The X-ray data seem to support our conclusion about the optical
classification of \4u, as many other  intermediate type Seyfert  galaxies 
have been found to show complex
absorption in the X-ray band, e.g. Mkn~6 (Feldmeier et al. 1999), 
NGC~4151 (Schurch \& Warwick 2003), Mkn~1152 (Quadrelli et al. 2003)  and
NGC~3227 (Gondoin et al. 2003).
In contrast, pure Type 1 objects have X-ray spectra that are
almost unobscured by cold absorption material (e.g. Nandra \& Pounds 1994;
Schartel et al. 1997; Reynolds 1997, Piconcelli et al. 2005).

According to the current Unified models for AGNs, Seyfert 1.5 are seen
at an intermediate inclination angle, and so it is possible that,
in the case of \4u,
our line-of-sight can intercept the outer (Compton-thin) layers of the torus 
and/or clouds in the so-called ``torus atmosphere'' (Feldmeier et
al. 1999, Lamer et al. 2003).
This provides a likely explanation for the presence of the complex
obscuration observed in this source. A deeper observation could be
useful to 
put useful constraints on ionization state and temporal behavior of
the absorbers.

Given the high column densities (\nhf~= 0.69$^{+0.26}_{-0.39}$
$\times$ 10$^{22}$ \cm2~and
\nhp~= 3.90$^{+2.15}_{-0.92}$  $\times$ 10$^{22}$ \cm2), the absorption dominates
the spectrum up to $\sim$ 3--4 keV. Above these energies 
the primary
X-ray power-law continuum starts to emerge with a $\Gamma$ = 1.54$^{+0.09}_{-0.10}$.
Beckmann et al. (2005)
have recently reported its significant detection up to $\sim$ 100 keV using {\it
  INTEGRAL} ISGRI spectroscopic data, measuring  a slope of
1.65$^{+0.02}_{-0.03}$, which is consistent with our result. 
Such a value of the photon index is slightly lower than the
average value $\Gamma \sim$ 1.9 reported for large samples of AGNs. However, our measurement
is not surprising since a flat
($\Gamma$ $\approx$ 1.5-1.6) continuum has been also observed
occasionally in other Seyfert 1.5 galaxies (e.g. Turner et al. 1999;
Gondoin et al. 2003; Schurch \& Warwick 2003).

\subsection{Properties of the Fe K$\alpha$ emission}

The present  \xmm~observation has revealed that the Fe K$\alpha$
fluorescence emission in \4u~is broad. A model with a 
relativistic (i.e. Kerr or Scharwzschild) profile (models C
and D, respectively) successfully reproduce  the data.

The discovery of such a spectral feature is very interesting since
 it may be interpreted as
the result of the emission from dense
matter in the innermost region of the accretion disk where effects
such as Doppler and gravitational broadening operate and give rise
to the typical asymmetric double-horned/skewed profile (Fabian et
al. 2000; Martocchia, Karas \& Matt 2000).
This should make possible to trace the matter distribution 
down to the last stable orbit.

Previous works based on {\it ASCA} data (e.g. Nandra et al. 1997, but
see also Lubinski \& Zdziarski 2001) 
report the common occurrence
of Fe K$\alpha$ lines with a broad profile in the X--ray spectrum of
most bright Seyfert galaxies, while in the AGNs
targeted by \xmm~and \chandra~the iron line has been
observed as a ubiquitous narrow feature at $\approx$6.4 keV 
(Reeves 2003; Yaqoob \& Padmanabhan 2004; Bianchi et al. 2004; 
Jim\'enez-Bailon et al. 2005). 
Only a small  number of objects show  an emission line with 
a  broadened, asymmetric
profile (Fabian \& Miniutti 2005
for a review) often superimposed on the narrow component (Reeves et
al. 2001; Balestra et
al. 2004).
The line detected in \4u~is one of the clearest examples of
  relativistic lines found so far. It is worth noting that
the feature shows a broad profile
independently from model fits applied to the underlying continuum.

The values of the line parameters derived from the spectral analysis of
the \xmm~observation suggest that in the case of \4u~the
fluorescence takes place at few gravitational radii from
the BH  with the inner radius
of the iron emitting region located at  $R_{\rm in}$ $\sim$ 10 R$_g$.
% This may explain the fact that both a {\tt laor} and a {\tt diskline}
% line profile  fit well the data. Furthermore 
Such a value of $R_{\rm in}$ does not allow us to estimate
the spin value $a$ of the BH.
The radius of the marginally stable orbit of a non-rotating (i.e. $a$ =
0) BH is 6
R$_g$, while in the case of a Kerr spinning ($a$ $>$ 0) BH it decrease down to 1.23
R$_g$: the line in \4u~is therefore not as extremely broad and
redshifted as those found in MCG--6-30-15 (Wilms et
al. 2001) and NGC 3516 (Turner et al. 2002),
 which imply a Fe
line-emitting disk extending down to the last stable orbit of the BH.

There are three straightforward explanations for 
the non-occurrence of Fe K$\alpha$ fluorescence
in the very immediate vicinity of the BH: ($i$)  the disk
does not extend below $\sim$ 10 R$_g$, ($ii$) the accretion flow very close
to the event horizon is  highly ionized and, therefore, the Fe is
completely stripped; and/or ($iii$) there are no X-ray emitting
active regions at \simlt~10 R$_g$ to properly 
illuminate the disk surface.

 However, it is worth noting that the spectral complexity, along with
the limited energy band, prevented us from a
fully and reliably understanding the underlying continuum. While
the existence of a relativistically broadened profile is robust
enough, the precise determination of the emitting disk parameters
must wait for future, broader band observations.

%%%%%%%%%%%%%%%%%%%%%%%%%%%%%%%%%%%%%%%%%%%%%%%
%%%%%%%%%%%%%%%%%%%%%%%%%%%%%%%%%%%%%%%%%%%%%%%%%
\subsection{The emission features around 5 keV}

\xmm~spectra of a handful of bright Seyfert galaxies have revealed the
presence of narrow emission lines in the energy band $\sim$ 5--7 keV (e.g. Turner
et al. 2002; Guainazzi 2003; Porquet et al. 2004; Turner et al. 2004;
Della Ceca et al 2005). 
We found  two examples of these features in the
\xmm~spectrum of \4u~at 4.90$\pm$0.04 (at 95.5\% c.l.) and
5.31$\pm$0.05 keV (at 98.5\% c.l.).

Some scenarios have been proposed recently to explain such lines
in terms of iron emission invoking a complex geometry 
of the Fe K$\alpha$--emitting region. 
If the X-ray corona is concentrated in a number of small
active regions (the so-called {\it flares}) that illuminate the
underlying disk (Haardt et al. 1994; Merloni \& Fabian 2001, Czerny et al. 2004),
Fe fluorescent emission lines can emerge from the 
localized  ``hotspots'' occurring on the accretion disk surface.
Dov{\v c}iak et al. (2004) present the detailed Fe
K$\alpha$ line profiles as a function of the orbital phase of the
hotspot and its radial distance from the BH.  The centroid of
emission line is then redshifted  due to the joint action of the
Doppler and gravitational energy shift.  The  few highly-redshifted emission
features observed around 5--6 keV so far were found to be
narrow. This leads to the belief that the emitting region is small
(i.e. a spot of a few $R_g$s) and that the irradiation lasts only for a
portion of the complete orbit.  

%Since the line originates from a
%rotating disk, the expected line profile is double--peaked: however,
%due to the limited spectral resolution/sensitivity of the current
%X--ray detectors, the observed features appear narrow and they are
%interpreted as the blue-horn (i.e. the most powerful one) of such a
%diskline arising from an orbiting spot (Dov{\v c}iak et al. 2004; 
%Pech\'acek et al. 2005). 
%This could also be the case for the features detected in 
%the spectrum of \4u. 
The detection of these lines (even if the
significance of the line at $\sim$ 4.9 keV is quite marginal)
would imply Fe K emission  close to the marginally stable orbit consistent
with an $R_{\rm in}$ \simgt~10 R$_g$ derived for the broad
line. This suggests the presence of different active regions, i.e.
an extended, long-lived corona uniformly illuminating the disk at
\simgt~10 R$_g$,
to give rise to the broad Fe feature, plus some flares that generate
hotspots in the innermost part of the accretion disk, giving rise 
to the narrow-line features around 5 keV.

Turner et al. (2002; 2004) 
propose an alternative scenario for the origin of
these lines based on a (mildly-) relativistic outflow.
In particular, they invoke the presence of a blob of 
Fe--emitting material ejected from the nucleus
to account for the significant shift in the energy of
the line observed over a few of tens kiloseconds and the
``transient'' behavior of the spectral line.
Assuming an origin in neutral iron, the line detected at 5.2(4.8) keV
in \4u~implies a velocity of $\sim$ 60,000 ($\sim$ 90,000) km/s of the
emitting material.

\section{Summary}

We have studied the optical and X--ray spectra of \4u, and our main
results can be summarized as follows:
\begin{itemize}

\item~\4u~is a very bright X--ray source with a \fhx~= 3.6 $\times$
  10$^{-11}$ \cgs. On the basis of the
  optical data 
we measured a redshift of $z$ = 0.012$\pm$0.001, which implies a \lum~$\sim$ 1.5 $\times$
10$^{43}$ \ergs, and we proposed to classify \4u~as an intermediate
  type Seyfert galaxy. In particular, 
since the \hbeta~line was not detected, 
we suggested a Seyfert 1.5 classification 
using the  ratio $I$(\nhalpha)/$I$(\bhalpha).
However, future, better-quality optical observations would allow us to obtain a more
  accurate constraint on the Seyfert type.\\
\item~Once corrected for the large Galactic extinction (\nhgal~$\sim$
  10$^{22}$ \cm2), \4u~still showed a heavily absorbed spectrum. 
The \xmm~data were best fitted 
by a power-law continuum with a slope of $\sim$ 1.55 
modified by two neutral absorption components (\nhf~$\sim$ 10$^{22}$ \cm2 and
  \nhp~$\sim$ 4 $\times$ 10$^{22}$
\cm2), the latter covering only the 50\% of the primary X--ray source.
X-ray data support the classification of \4u~as an intermediate type
Seyfert galaxy.\\
\item~\4u~exhibits a broad and skewed Fe K$\alpha$ line. This finding 
allows the small sample
 of broad Fe K lines observed with
\xmm~so far to be increased. 
This spectral feature is a signature for the presence of an
accretion disk that is able to reprocess the primary continuum  at
a few gravitational radii ($\sim$~10) from the marginally stable orbit around the
black hole. The line does not show any extreme
relativistically-broadened profile, so it can be fitted by a
Schwarzschild black hole model. No conclusion can be drawn about the spin properties of the
BH, if any. The energy of the line is consistent with nearly neutral iron.
We found that an additional,  narrow contribution to the Fe K$\alpha$
emission  (e.g. from a distant
reflector) is not required by the data.
Given the brightness of \4u, future high signal-to-noise observations carried out by \xmm~or {\it
Suzaku} could be helpful for determining the line parameters on firmer
statistical grounds.\\
\item~Our analysis has revealed the presence of two narrow line-like
  features at $\sim$ 4.9 and $\sim$ 5.3 keV, significant at the 95.5\% and
  98.5\% confidence levels, respectively.  Recently, similar features were 
also observed in the X-ray spectra of a handful of bright Seyfert
  galaxies. Their nature remains, however,  puzzling. A possible explanation
 has been proposed by Dov{\v c}iak et al. (2004) in terms of
  highly-redshifted iron line produced in a localized hot spot on the
  surface of the accretion disk very close to the marginally stable orbit of the BH.
The simultaneous presence of a broad line in the spectrum of
  \4u~reinforces this hypothesis thanks to the direct evidence of an
  Fe-emitting,  relativistic accretion disk in this source.
 Long uninterrupted \xmm~observations could lead to useful insights into the
nature of these
emission features: in particular, the detection of variability in the
shape and  energy  centroid of a line is crucial to accurately
determining where it originates and how it evolves (e.g. Iwasawa,
Miniutti \& Fabian 2004). 
\end{itemize}
%%%%%%%%%%%%%%%%%%%%%%%%%%%%%%%%%%%%%%%%%%%%%%%%%%%%%%%%%%%%%%%%%%%%%%%%%%%%%%%%%%%
\begin{acknowledgements}

The authors wish to thank the referee for a prompt report that led to an improved paper.
We are grateful to the staff members of the \xmm~Science
Operations Center for their help. We also thank Martin Stuhlinger
(ESAC) for helpful discussions related with the {\it EPIC} calibrations status. 
EP acknowledges the financial support of INSA (Spain).

\end{acknowledgements}
%%%%%%%%%%%%%%%%%%%%%%%%%%%%%%%%%%%%%%%%%%%%%%%%%%%%%%%%%%%%%%%%%%%%%%%%%%%%%%%%%%%

\end{document}